\documentclass[nofootinbib,showpacs,amsmath,twocolumn,showkeys,10pt,aps,pra,longbibliography,floatfix]{revtex4-2}
\usepackage[T1]{fontenc}
\usepackage{inputenc}
\usepackage[margin=1in]{geometry}
\usepackage{color,times}
\usepackage{array}
\usepackage{amsmath}
\usepackage{stackrel}
\usepackage{graphicx}
\usepackage{appendix}
\usepackage{mathrsfs}
\usepackage{amsfonts}
\usepackage{appendix}
\usepackage{color}
\usepackage{comment}
\usepackage{soul}
\soulregister\cite7
\soulregister\ref7
\usepackage{xcolor}
\usepackage [english]{babel}
\usepackage [autostyle, english = american]{csquotes}
\MakeInnerQuote{"}
\definecolor{med-blue}{RGB}{25,25,112} 
\allowdisplaybreaks

\newcommand{\ket}[1]{\vert{#1}\rangle}
\newcommand{\bra}[1]{\langle{#1}\vert}
\newcommand{\outpr}[2]{\vert{#1}\rangle\langle{#2}\vert}

\newcommand{\expv}[1]{\langle{#1}\rangle}

\begin{document}
\title{High fidelity single-qubit quantum state tomography of electron-$^{14}$N nuclear hybrid spin register in diamond using Rabi oscillations}
\author{Abhishek Shukla$^{1,2,a}$}\email{abhishek.shukla@uhasselt.be} \author{Boo Carmans$^{1,a}$} \author{Michael Petrov$^{1,a}$} \author{Daan Vrancken$^{1}$}\thanks{Currently: centre for Molecular Modeling (CMM), Ghent University, Technologiepark 46, 9052 Zwijnaarde, Belgium.} \author{Milos Nesladek$^{1,2}$}\email{milos.nesladek@uhasselt.be}
\affiliation{$^a$Authors contributed equally}
\affiliation{
$^{1}$Q-Lab, Hasselt University, Wetenschapspark 1, 3590 Diepenbeek, Belgium.}
\affiliation{$^{2}$IMOMEC division IMEC, Wetenschapspark 1, 3590 Diepenbeek, Belgium.}

\begin{abstract}
   We report on a novel quantum state characterisation method, which we call Rabi-based Quantum State Tomography (RQST), that we have validated on single-qubit quantum states, in particular on the electron and nuclear spins of a single nitrogen-vacancy (NV) centre in diamond, demonstrating high fidelities. The difference of RQST with conventional tomography methods is in the implementation of rotation operators and construction of the density matrix from the measured data sets. We demonstrate efficient quantum state control of the electron spin at room temperature with an average fidelity of 0.995 over more than 40 measurements on different states on the Bloch sphere. Also, we apply the methodology to the dark NV nuclear spin state. The state is read via the electron spin using the C-NOT two-qubit entanglement gate and demonstrates fidelities of the same order.  

\end{abstract}
\maketitle
\section{Introduction}\label{Intro}
The importance of quantum computation (QC) is growing with recent demonstrations towards quantum advantage \cite{arute2019quantum, zhong2020quantum,riste2017demonstration,morvan2023phase} as well as with approaching applications ranging from science and technology to daily life.\cite{andreas2021industry} This has motivated researchers to search for new methodologies for precise quantum state measurements, either with readout protocols or with post-processing.\cite{Zhang23,Anjou14,Rong2015} These methodologies are typically known as quantum state tomography (QST), i.e. the method for characterisation of a quantum state, and quantum process tomography (QPT), i.e. the method for characterisation of quantum gates.\cite{nielsen2010quantum} They are at the heart of quantum engineering and scaling up to large qubit systems.\\
Various quantum computing architectures,\cite{nielsen2010quantum} have shown an encouraging progress in achieving intermediate-scale quantum registers and developing a noisy intermediate-scale quantum computer. For example, the silicon-vacancy (SiV) centre appears as a candidate for QC realisation.\cite{neu2013low, Metsch} However, group-IV vacancy complexes need to operate at low temperatures of about 100 mK up to the 5 K range.\cite{Ivady} Among different quantum systems, the Nitrogen Vacancy (NV) diamond solid-state spin-based architecture, which has also been originally proposed as suitable for QC due to its compatibility with microelectronic systems, still needs to demonstrate its potential.\cite{Wrachtrup06} An advantage of the NV centre is the record coherence time at room temperature, in the ms range.\cite{Balasubramanian, DOHERTY2013,stolze2008quantum} Beside the high coherence time, the qualities that make the NV centre stand out are room temperature operation, high-fidelity optical initialisation and simple optical detection of magnetic resonances (ODMR) for the electron spin state.\\
One of the challenges in realizing solid-state spin-based QC is scalability: it is necessary to entangle a large number of qubits for QC operation and to read them out deterministically. Particularly, this is a problem for detecting dipole-dipole coupled NV electron spins by ODMR, which require spatially resolved readout on the scale of 10-20 nm. This important issue is addressed in ongoing works by trying to couple close NV spins using different pathways, like optically mediated coupling based on spin-photon interface or exploiting dipole-dipole coupling and its electrical readout.\cite{siyushev2019photoelectrical,ruf2021quantum}\\
Another important requirement of a quantum processor is fault tolerance, which can be achieved by applying error correction codes. Due to their weak coupling with the environment and a robust control, solid-state spin systems are arising as a suitable candidate for fault-tolerant quantum computation.\cite{hanson2008coherent} This makes NV an eligible candidate for quantum computation, specifically desktop QC, if the scalability bottleneck can be broken.\cite{colorcenterInDiamonds}\\ 
We design a novel quantum state tomography method based on Rabi experiments combined with single-qubit spin operations and entangling spin gates used for the preparation and read-out of the dark spin state. In the case of a single qubit, we replace the unitary operators, which are used to read the unobservable terms, with an array of operators that comprise a Rabi cycle. In particular, instead of $X_{90}$ and $Y_{90}$ operations used in other QST methods,\cite{Chuang,Zhang23} we perform multiple operations like $X_{5}$, $X_{10}$, $X_{15}$, etc. We call this method Rabi Quantum State Tomography (RQST). Though, this technique has been suggested before,\cite{Yang16} it had not been fully developed or properly characterized. We present two variants of RQST: Rabi amplitude quantum state tomography (RAQST) and Rabi phase quantum state tomography (RPQST) depending on whether amplitude or phase information of the Rabi oscillation is used for extracting quantum state parameters. Through numerical simulation we show that using RQST can be advantageous at certain conditions.\\ 
Further, we demonstrate the application of the proposed method on the readout of the electron and nuclear spins of an NV centre. An NV centre is a point defect in the diamond lattice consisting of a substitutional nitrogen atom and a neighbouring vacancy. It can be neutral (NV$^0$), negatively (NV$^-$) or positively (NV$^+$) charged. The NV$^-$ has a spin-triplet ground state since two of its electrons are unpaired. NV$^-$ (further referred to as "NV" in this article) possesses an electron spin and a $^{14}$N nuclear spin, both of which are spin 1 systems forming a 9-dimensional Hilbert space of electron-$^{14}$N (E-N) composite spin systems.\cite{DOHERTY2013} Exploiting universal control and careful readout of only the desired subspace, it is possible to prepare an arbitrary quantum state over the reduced 4-dimensional Hilbert space corresponding to two two-level spin subsystems, of which the electronic spin is used as a probe by measuring optically detected magnetic resonance. The E-N spin register system, and the Hamiltonian along with experimental details, are described in Supplementary Material \cite{Supp} and we demonstrate the method on the NV electron spin alone as well as on the nuclear spin of the E-N system using NV electron spin as readout ancilla via two-qubit gates.\\
Taking the initialisation procedure from \cite{chakraborty2017polarizing, Neumann} as a base, we modify and optimize the spin drive and readout sequence that allows us to initialize both electron and nuclear spins to $m_{s}=0,m_{I}=0$ indicated as the state $\ket{0,0}$. Further, we prepare an arbitrary initial single-qubit state on the nuclear spin and entangle it with the NV electron spin-qubit. This is particularly important as the nuclear spin cannot be directly read optically. Such an entanglement allows us to read nuclear spin state population via two-qubit gates.\cite{Neumann}\\
In this work, we focus on the characterisation of pure states, as opposed to mixed states, owing to the interest in quantum computing applications. In practice, due to the effect of decoherence and systematic errors, or as a matter of interest, mixed states may arise in the course of general quantum dynamics and hence it may also be interesting to have a characterisation technique for them, i.e. mixed state QST.  We provide a short discussion on state mixing in Supplementary Material \cite{Supp}, a detailed study is the scope of the upcoming future work.\\

\section{Rabi quantum state tomography. Description and comparison.} \label{theory}
The most general single-qubit quantum state $\ket{\psi}$ can be described in the Pauli operator basis $\{\mathbb{I},\sigma_{x},\sigma_{y},\sigma_{z}\}$ as
\begin{eqnarray}
       \rho &=& \frac{1}{2}(\mathbb{I}+\expv{\sigma_{x}}\sigma_{x}+\expv{\sigma_{y}}\sigma_{y}+\expv{\sigma_{z}}\sigma_{z})\\ \nonumber
        &=& \frac{1}{2}(\mathbb{I}+n_{x}\sigma_{x}+n_{y}\sigma_{y}+n_{z}\sigma_{z})
        \label{rho}
\end{eqnarray}
Here, $\rho$ is the density matrix of the $\ket{\psi}$ state vector, $n_{x}$, $n_{y}$, and $n_{z}$ are the coordinates of the Bloch vector tip, which may possibly be obtained through classical projection measurements. The characterisation of the quantum state is equivalent to calculating the coordinates of the point coinciding with the tip of the Bloch sphere, i.e., $n_{x}$, $n_{y}$, and $n_{z}$. In the case of the NV centre electron spin, only the diagonal terms of the density matrix are directly observable (through the intensity of photoluminescence (PL) emitted by the NV under laser illumination).\\
In, e.g., a projection-based QST method \cite{Chuang,Zhang23}, to measure the off-diagonal terms one needs to apply unitary operators to transfer them to the diagonal terms.
\begin{eqnarray}
\nonumber n_x &=& \text{diag}(U_{1} \rho' U_{1}^{\dagger})\\
n_y &=& \text{diag}(U_{2} \rho' U_{2}^{\dagger})\\
\nonumber n_z &=& \text{diag}(\rho')
 \label{StndrdQST}
\end{eqnarray}
where $\rho'$ is a traceless part of the density matrix, diag is the diagonal part of the matrix, and $U_1$ and $U_2$ are the unitary operators, which for the single qubit case are simply the $X_{90}$ and $Y_{90}$ operations ($\pi/2$-pulses).\cite{Zhang23}\\
In our approach, we use two Rabi experiments combined with spin gates to determine the quantum state. As mentioned above, there are two ways to extract $n_x$, $n_y$, $n_z$ from Rabi experiments. In particular, the RAQST method uses the amplitudes of the Rabi oscillations for the quantum state reconstruction and the RPQST method uses only the phases of the X and Y Rabi oscillations. We explain first the RAQST in more detail.\\
\begin{figure}[htbp]
    \centering
    \includegraphics[width=\linewidth]{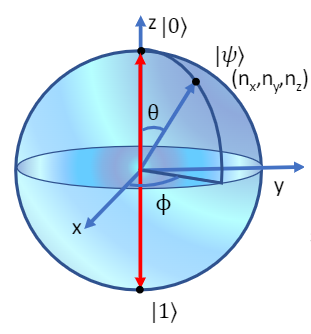}
    \caption{Bloch sphere representation of an unknown state $\ket{\psi}$ with Cartesian coordinates  $n_{x}, n_{y}, n_{z}$ and polar coordinates $\theta$ and $\phi$. The state depicted is pure and hence represented by the Bloch vector with the tip of the vector on the surface of the Bloch sphere, while a mixed state would be represented by the Bloch vector with the tip inside the Bloch sphere.}
    \label{fig1:Blochsphere}
\end{figure}
The spherical geometry of the Bloch sphere gives rise to the following relations between state coordinates
      \begin{eqnarray}
     \nonumber n_{x}^{2} &=& 1-\left(\frac{A_{x}}{A_\text{ref}}\right)^{2},\label{eq4:coordinatecalculation}\\ 
      n_{y}^{2} &=& 1-\left(\frac{A_{y}}{A_\text{ref}}\right)^{2},\\ \nonumber
      n_{z}^{2} &=& \left(\frac{A_{x}}{A_\text{ref}}\right)^{2} + \left(\frac{A_{y}}{A_\text{ref}}\right)^{2}-1.
      \end{eqnarray}
 Here $A_x$, $A_y$ are the amplitudes of the Rabi oscillations around the x- and y-axis respectively. $A_{\text{ref}}$ is the amplitude of the Rabi oscillation performed on an eigenstate. The RAQST method requires such a reference experiment to learn the largest Rabi amplitude $A_\text{ref}$, as the Bloch sphere is normalized. However, (\ref{eq4:coordinatecalculation}) does not uniquely specify the coordinates but rather narrows down possibilities to 8 points on the Bloch sphere, as seen in figure \ref{fig2:octants}b. Additional information has to be obtained from the phase of the Rabi oscillations (see Supplementary Material \cite{Supp}). After considering this phase information we can uniquely determine the coordinates $\{n_{x},n_{y},n_{z}\}$.\\
 
\begin{figure}[htbp]
    \centering
    \includegraphics[width=\linewidth]{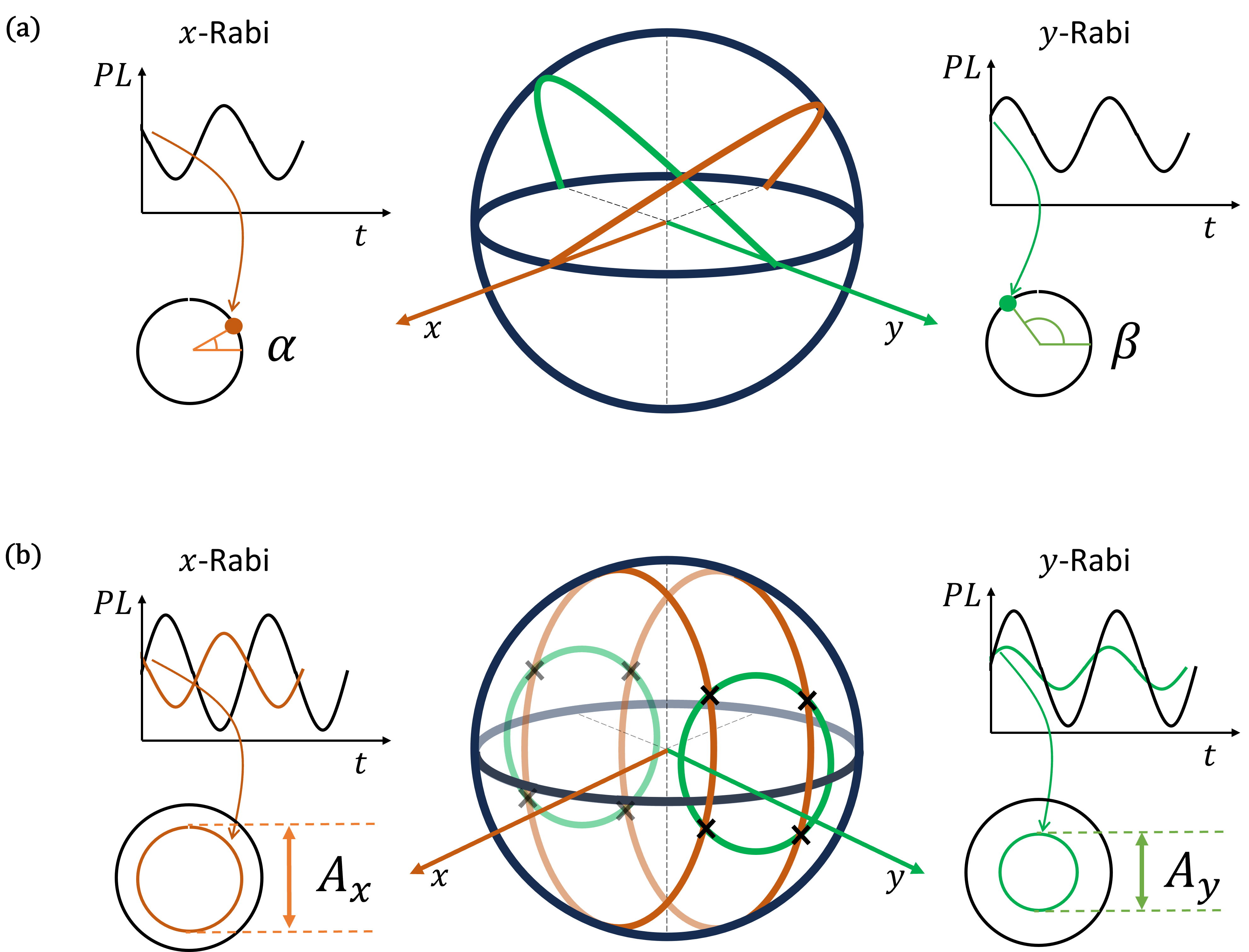}
    \caption{Schematic representation of RQST. In the centre is the Bloch sphere, a single point on which represents a superposition state of two-level system unequivocally. On the left and right are the x- and y-Rabi measurements, which will allow to find this point. \textbf{(a)} Phase Rabi QST. The x-Rabi measurement narrows down the location of this point to a single (brown) line, where the phase angle is equal to $\alpha$. The intersection points with the similar (green) line from a $90^{\circ}$ shifted Rabi measurement (y-Rabi) with phase angle $\beta$ is the result of the tomography. \textbf{(b)} Amplitude Rabi QST. The x-Rabi measurement defines two circles on the surface of the Bloch sphere with Rabi amplitude equal to $A_x$ (brown) and the y-Rabi does the same for amplitude $A_y$ (green). The state vector tip is then limited to the eight crossing points on the Bloch sphere. This can be reduced to a single point, after considering the phase information, thus uniquely determining the state coordinates as discussed in the text. The black traces on the left and right represent Rabi rotations of an eigenstate and their amplitude is used as the Bloch sphere has unit radius.}
    \label{fig2:octants}  
\end{figure}
Alternatively, we can describe the pure state $\ket{\psi}$ as
\begin{equation}
        \ket{\psi} = \cos{\frac{\theta}{2}}\ket{0}+e^{i \phi}\sin{\frac{\theta}{2}}\ket{1}.
        \label{psi}
\end{equation}
where quantities $\theta$ and $\phi$ are respectively the polar and the azimuthal angle in the Bloch sphere representation (Fig. \ref{fig1:Blochsphere}). The polar angle $\theta$ and the azimuthal angle $\phi$ used in (\ref{psi}) are related with the coordinates $n_{x}$, $n_{y}$, and $n_{z}$ by the following formulae:
       \begin{align}
       \tan\left(\frac\pi 2 - \theta \right) &= \frac{\sqrt{n_{x}^2+n_{y}^2}}{n_{z}} \\
       \tan\phi &= \frac{n_x}{n_y}
       \label{eq5:thetaAndPhi}
       \end{align}   
 To summarize, using the Rabi amplitudes ($A_x$ and $A_y$) of x-Rabi and y-Rabi measurements of an unknown state together with the reference amplitude ($A_\text{ref}$) allows us to calculate the Cartesian and the Euler coordinates of the tip of the quantum state vector on the Bloch sphere $n{_x}$, $n{_y}$, $n_{z}$, $\theta$ and $\phi$.\\
 Figure \ref{fig2:octants}a introduces angles $\alpha$ and $\beta$, which are extracted by fitting the Rabi oscillations with PL $= \sin{(\omega t +\alpha)}$ (fig. \ref{fig2:octants}a). In principle, one could do such a fitting with $\omega$ as a free parameter, however, in practice, to achieve fidelity comparable with the projection-based method, it is necessary to know $\omega$ in advance from a long reference measurement.  The angles $\theta$ and $\phi$ of the state vector are related with the phases of the Rabi oscillations $\alpha$ and $\beta$ as follows:
\begin{eqnarray}
\tan\left(\frac\pi 2 - \theta \right) = \frac{\tan{\alpha}}{\cos{\phi}}\\
\tan\phi = \frac{\tan{\beta}}{\tan{\alpha}}
\label{phh}
\end{eqnarray}

\begin{figure}[htbp]
    \centering
    \includegraphics[width=\linewidth]{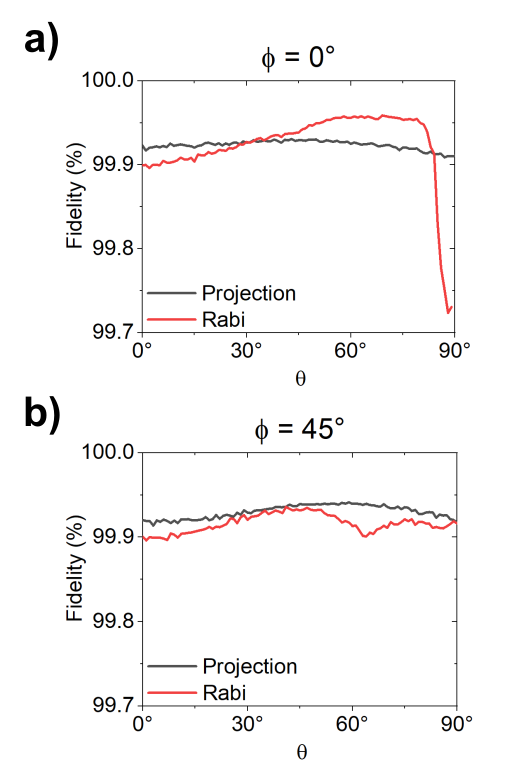}
    \caption{Results of the simulation. Fidelity of spin-state coordinates obtained by projection and Rabi QST for \textbf{(a)} $\phi = 0^{\circ}$ and \textbf{(b)} $\phi = 45^{\circ}$. The performance of Rabi tomography lies close to projection-based tomography, for certain angles surpassing it.}
    \label{sim1}
\end{figure}

\begin{figure}[htbp]
    \centering
    \includegraphics[width=\linewidth]{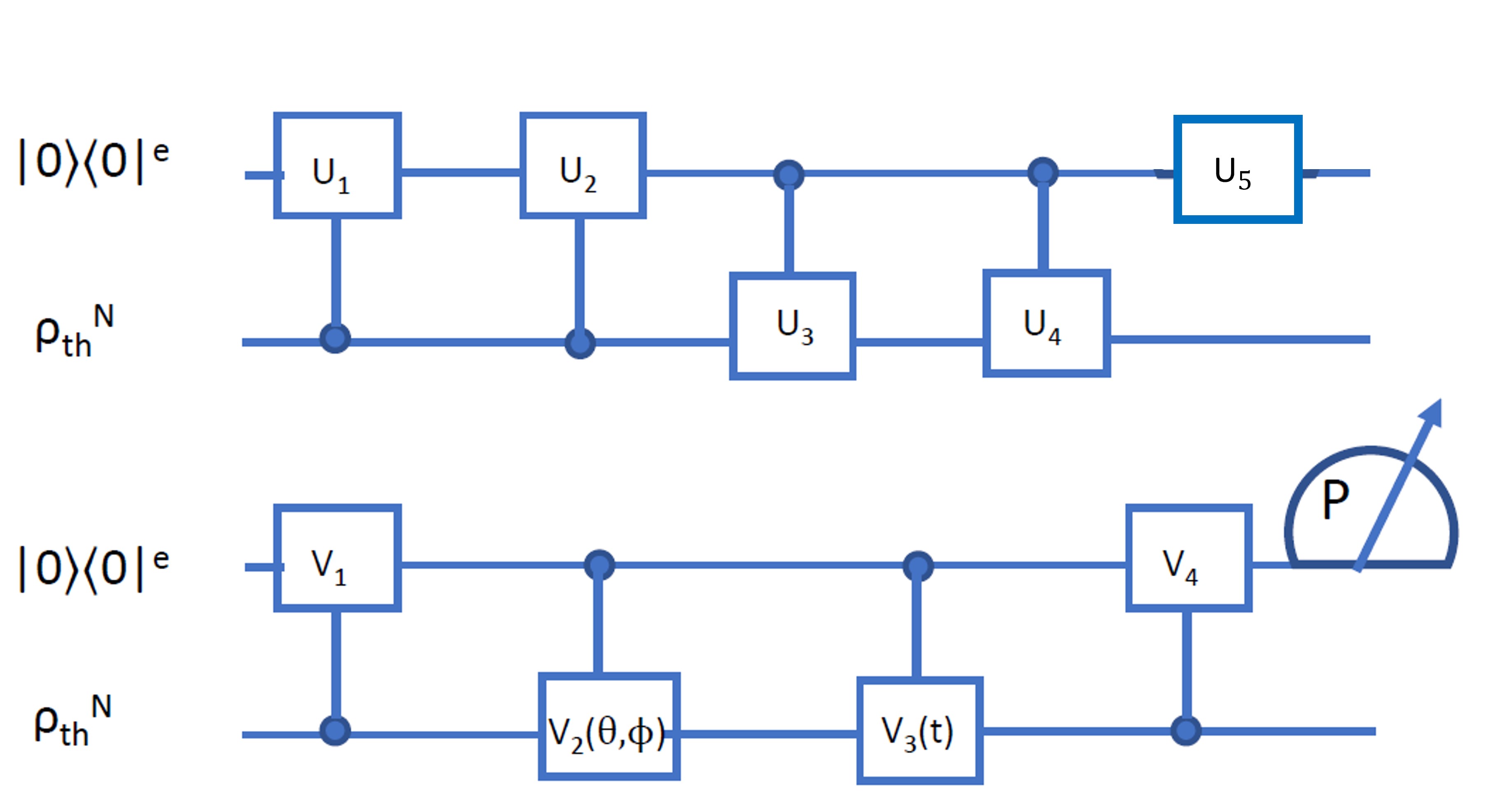}
    \vspace{0.5cm}
    \caption{\textbf{(a)} Quantum circuit for initialisation of electron-nuclear spin two-qubit hybrid register to state $\ket{0,0}$. \textbf{(b)} Preparation of nuclear spin qubit to state $\ket{\psi}$ and operation for performing Rabi oscillation. The unitary operators U and V are described in computational basis in the text.} \label{circuit}
\end{figure}

We reconstruct the experimental density matrix $\rho_\text{exp}$ such that $\rho_\text{exp}= \ket{\psi_\text{exp}}\bra{\psi_\text{exp}}$ with $\ket{\psi_\text{exp}}$ the experimentally determined state on the Bloch sphere. From this, we calculate the fidelities
\begin{equation}
F = \frac{\text{Tr}(\rho_\text{th} \rho_\text{exp})}{\sqrt{\text{Tr}(\rho_\text{th} ^2)\text{Tr}(\rho_\text{exp}^2)}}\,. \label{eq:fidelity}
\end{equation}
where $\rho_\text{th}$ is the density matrix corresponding to the prepared state $\ket{\psi_\text{th}}$.\\
Using the defined equations for RQST, we perform a numerical simulation to compare our proposed method to the projection-based tomography \cite{Zhang23} in presence of uniform white noise (fig. \ref{sim1}). We see from the results, that for certain $\theta$ and $\phi$ RQST performs better than projection-based QST. However, we note that RQST performs weaker when the measured spin state is close to the equator of the Bloch sphere (fig \ref{sim1}a). Presented are the examples for $\phi = 0^{\circ}$ and $\phi = 45^{\circ}$,
for more examples and details on the simulation, see Supplementary Material \cite{Supp}.\\
From the both RAQST and RPQST methodologies,  the RPQST has the advantage of being robust against sample drift.  In particular, when measuring on a single defect, it can drift out of focus during one of the Rabi measurements, altering its amplitude. The phase of the Rabi measurements is less affected by such drifts. On the other hand, monitoring the amplitude in RAQST allows to check for the presence of this sample drift and allows also to analyse the state mixing, for example due to decoherence, during the state evolution.\\
The presented methodology is however general and allows for the fidelity evaluation of practically arbitrary type of qubits. In particular for our measurements, in addition to the NV electron spin, we demonstrate the new QST methodology on the NV nuclear spin. The nuclear spin coherence time is significantly higher than for the electron spin, which is crucial for realising quantum memories \cite{DOHERTY2013}. However, unlike the NV electron spin, the nuclear spin is not directly observable, i.e. has no visible optically related transition, therefore, the necessary pulse protocols are involved and require electron-nuclear gates to be applied. Here, the nuclear spin is manipulated using radio-frequency (RF) driving and read out using an ancillary electron spin exploiting entanglement, i.e. two-qubit gates. Taking this into account, though the proposed Rabi oscillation method targets only one spin qubit, in reality the QST fidelity inherently depends on the fidelity with which we execute the two-qubit gates.\\
In the following, we provide a quantum circuit describing the method for initialising the nuclear spin to state $\ket{0}$ (Fig. \ref{circuit}a)\cite{chakraborty2017polarizing} and a quantum circuit for our QST method (Fig. \ref{circuit}b). The method is formulated for the E-N system as a qutrit-qutrit system with a general state vector

\begin{equation}
    \ket{\psi}=\ket{a,b}
\end{equation}
where $a$ and $b$ are the states of the electron and nuclear spins, respectively, and $\{a,b\}$ $\in$ $[0,-1, 1]$. We also describe all the conditional operations as controlled gates in their operator form. 

The operators used in the circuit for initialisation of E-N to state $\ket{0,0}$ (Fig. \ref{circuit}a) are given as follows:
\newline
$U_{1}: \operatorname{I_{3}}\otimes \outpr{0}{0}+\operatorname{I_{3}} \otimes \outpr{1}{1}+(\outpr{1}{0}+\outpr{0}{1}+\outpr{-1}{-1})\otimes \ket{-1}\bra{-1}$\\
\newline
$U_{2}\colon \operatorname{I_{3}} \otimes\ket{0}\bra{0}+\operatorname{I_{3}}\otimes \ket{-1}\bra{-1}+(\ket{1}\bra{1}+\ket{-1}\bra{0}+\ket{0}\bra{-1})\otimes \ket{1}\bra{1}$\\
\newline
$U_{3}\colon \ket{-1}\bra{-1}\otimes \operatorname{I_{3}}+\ket{-1}\bra{-1}\otimes \operatorname{I_{3}}+\ket{1}\bra{1}\otimes(\ket{0}\bra{-1}+\ket{-1}\bra{0}+\ket{1}\bra{1})$\\
\newline
$U_{4}\colon \ket{1}\bra{1}\otimes \operatorname{I_{3}}+\ket{0}\bra{0}\otimes \operatorname{I_3}+\ket{-1}\bra{-1}\otimes(\ket{0}\bra{1}+\ket{1}\bra{0}+\ket{-1}\bra{-1})$\\
\newline
$U_{5} \colon (\ket0\bra0+\ket0\bra1+\ket0\bra{-1})\otimes\operatorname{I_{3}}$\\

Here, $U_1$ is the the $\pi$-pulse on the electron spin from $\ket0$ to $\ket1$ conditional on the nuclear spin being at $\ket{-1}$; $U_2$ is the $\pi$-pulse on the electron spin from $\ket0$ to $\ket{-1}$ conditional on the nuclear spin being at $\ket1$; $U_3$ is the $\pi$-pulse on the nuclear spin from $\ket0$ to $\ket{-1}$ conditional on the electron spin being at $\ket1$; $U_4$ is the $\pi$-pulse on the nuclear spin from $\ket0$ to $\ket1$ conditional on the electron spin being at $\ket{-1}$;  $U_5$ is the unconditional initialization of the electron spin to $\ket0$.\\
After the initialisation of E-N to state $\ket{0,0}$, figure \ref{circuit}b illustrates the preparation of the state to be tomographed followed by a definite phase Rabi experiment. Here, we first transfer electron spin from state $\ket{0}$ to $\ket{-1}$ conditional to the nuclear spin state $\ket{0}$ using operator \\ 
$V_{1}\colon (\ket{-1}\bra0+\ket{0}\bra{-1}+\ket1\bra1)\ket0\bra0+\operatorname{I_3} \otimes \ket{1}\bra{1}+\operatorname{I_3} \otimes \ket{-1}\bra{-1}$\\
We apply conditional gate operations for the preparation of the desired quantum state $\ket{\psi}$ on the nuclear spin and for executing nuclear Rabi operations. These two operations are limited to the nuclear spin subspace (i.e., $\ket{1}$, $\ket{0}$) of the electron spin subspace $\ket{-1}$ in the two-dimensional Hilbert space. The measurement is carried out in the $\ket{0}$ electron spin subspace. We can then consider the dynamics of the two-qubit register effectively limited to the subspace ($\ket{0,1},\ket{0,0},\ket{-1,1},\ket{-1,0}$) to describe the dynamics without loss of generality. After the application of the $V_{1}$ 
operator, the quantum system is in the state $\ket{-1,0}$. We then prepare a quantum state $\ket{\psi}$ with tilt angle $\theta$ and phase $\phi$ by using operator $V_{2}$ 
(Fig. \ref{circuit}b). The state now becomes of the form

\begin{equation}
\begin{split}
\ket{\psi_{1}}=\cos{\frac{\theta}{2}}\ket{-1,0}+e^{i\phi}\sin{\frac{\theta}{2}}\ket{-1,1}    
\end{split}
\end{equation}

 Furthermore, parametric gate $V_{3}(t)$ is used for applying the Rabi operation conditional to electron spin subspace $\ket{-1}$. Let's say, $\theta_{R}$ is the Rabi nutation angle at time instant $t$ then the state becomes
 \begin{equation}
 \begin{split}
\ket{\psi_{2}}= \{\cos{\frac{\theta}{2}}\cos{\frac{\theta_{R}}{2}}-i e^{i\phi}\sin{\frac{\theta}{2}}\sin{\frac{\theta_{R}}{2}}\}\ket{-1,0}\\
+\{-i\sin{\frac{\theta_{R}}{2}}\cos{\frac{\theta}{2}}+e^{i\phi}\sin{\frac{\theta}{2}}\cos{\frac{\theta_{R}}{2}}\}\ket{-1,1}
\end{split}
\label{StateAfterRabi}
\end{equation}
Ultimately, we entangle the nuclear and electron spin states by selectively flipping the electron spin state, in this case, the nuclear spin is in state $\ket{0}$. 
This is modelled as a controlled operation $V_{4}$, 
 with the nuclear spin state as a control and the electron spin state as a target. Now, the entangled state is 
 \begin{equation}
 \begin{split}
\ket{\psi_{3}}=\{\cos{\frac{\theta}{2}}\cos{\frac{\theta_{R}}{2}}-i e^{i\phi}\sin{\frac{\theta}{2}}\sin{\frac{\theta_{R}}{2}}\}\ket{0,0}\\
+\{-i\sin{\frac{\theta_{R}}{2}}\cos{\frac{\theta}{2}}+ e^{i\phi}\sin{\frac{\theta}{2}}\cos{\frac{\theta_{R}}{2}}\}\ket{-1,1}
\end{split}
\label{FinalState}
\end{equation}
 The nuclear spin population is then estimated through optical readout of electron spin population. Operations $V_{2}$ 
and $V_{3}$ 
are defined as follows:\newline
$V_{2}$$
: \outpr{0}{0} \otimes \operatorname{I_{3}} +\outpr{1}{1} \otimes \operatorname{I_{3}} +\outpr{-1}{-1} \otimes e^{(-i \theta I_{\phi})}$, \newline where $I_{\phi}= \cos{\phi}I_{x}+\sin{\phi}I_{y}$.\newline
$V_{3}$$
: \outpr{0}{0} \otimes \operatorname{I_{3}} +\outpr{1}{1} \otimes \operatorname{I_{3}} +\outpr{-1}{-1} \otimes e^{(-i \theta(t) I_{\zeta})}$.\newline
The above $I_{x}, I_{y}$ are spin angular momentum operators for a two-level system.\\
\section{Experimental implementation of RQST on NV centre system}

In the following, we discuss the experimental implementation of the single-qubit RQST methods on both electron and nuclear spins, i.e., electron RQST and nuclear RQST of the E$-^{14}N$ hybrid spin system in the diamond NV centre.\\

The electron-nuclear system of the NV centre, without considering strain, is described by the Hamiltonian
\begin{equation}
    H = 2\pi \hbar (D S_{z}^2 + \gamma_{e}B_{o}S_{z}+ PI_{z}^2 + \gamma_{N}B_{o}I_{z}+ A S_{z}I_{z})
    \label{Hamiltonian}
\end{equation}
Here, D is the zero-field splitting (ZFS), $\gamma_{e}B_{o}$ is the Zeeman splitting of electron spins, P is the quadrupolar splitting, $\gamma_{N}B_{o}$ is the Zeeman splitting of nuclear spins, A is the longitudinal part of the hyperfine coupling between electron and $^{14}$N nuclear spin. The three vibronic levels of the electron spin-1 system are the lowest vibronic levels out of which a two-level system can be derived, say $m_{s}=0$ for the ground vibronic state and $m_{s}=1$ for the excited vibronic state. The optical response of the NV centre depends on the respective population of these states, which allows for the readout operation. The qubit subspace $\{m_{I}=0,m_{I}=1\}$ from the nuclear spin state space can be derived analogously.  These two pairs of states form the computational space and the dynamics of the derived two-qubit system is defined in this computational space. Quantum mechanically, the populations of the two levels of the computational space are comprised of expectation values of the $\sigma_{z}$ operator. In contrast to the electron spin, the nuclear spin has no readable optical signature (it is a dark spin). Thus, $\sigma_{z}$ of the electron spin is the directly observable term, which we determine through Rabi measurements. In the Bloch sphere picture the population corresponds to the projection of the Bloch vector on the computational basis vectors.\\

We now describe the details depending on the spin type in the following two subsections. 

\subsection{QST implementation on single electron spin}
The viability of the presented QST methodology was first tested on the NV electron spin system as electronic spin transitions can be driven in nanosecond timescales, allowing for faster measurements and thus better statistics compared to measurements on the NV nuclear spin. 
The experimental implementation of QST on the NV electron spin is rather straightforward. We first initialize the electron spin using a green laser pulse of duration 3 $\mu$s, followed by a resonant Rabi pulse of fixed amplitude (1.25 MHz) and a duration interval of (0 to 3$\mu$s) with 100 sampling points. More details are given in Supplementary Material \cite{Supp}. It is important to note that the optical initialisation of the electron spin is limited by conversion to an optically undetectable NV$^{0}$ charge state. A remedy to this is in principle possible and can be performed using an orange laser for back conversion of NV$^{0}$ charge state to NV$^{-}$.\cite{Fedor} It should be noted that even in that case the polarisation of the electron spin to its $\ket0$ is not 100$\%$. For example in \cite{Zhang23} a time dependent photon counting was used as an alternative method to spin-projection based QST. In that case, fidelities of about 0.99 were reported, also not considering the full $\ket0$ initialization. Here, similarly to the argumentation in \cite{Zhang23} we work with maximal polarisation obtained by the laser pulse. We further discuss the influence of the state mixedness on the QST results in Supplementary Material \cite{Supp}. 

\begin{figure*}[t]
    \centering
    \includegraphics[width=8cm]{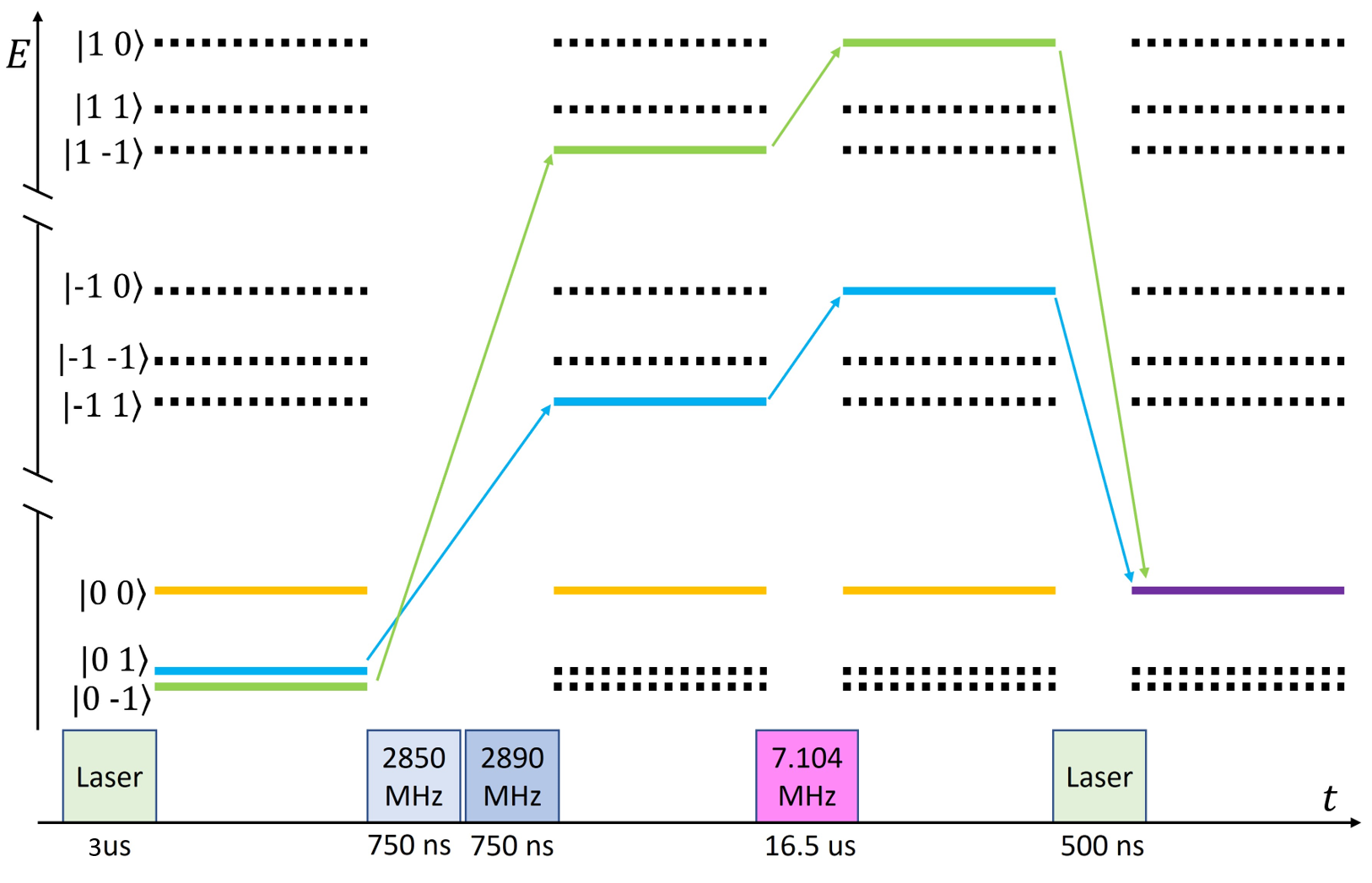}
    \label{nvtom}
    \includegraphics[width=8cm]{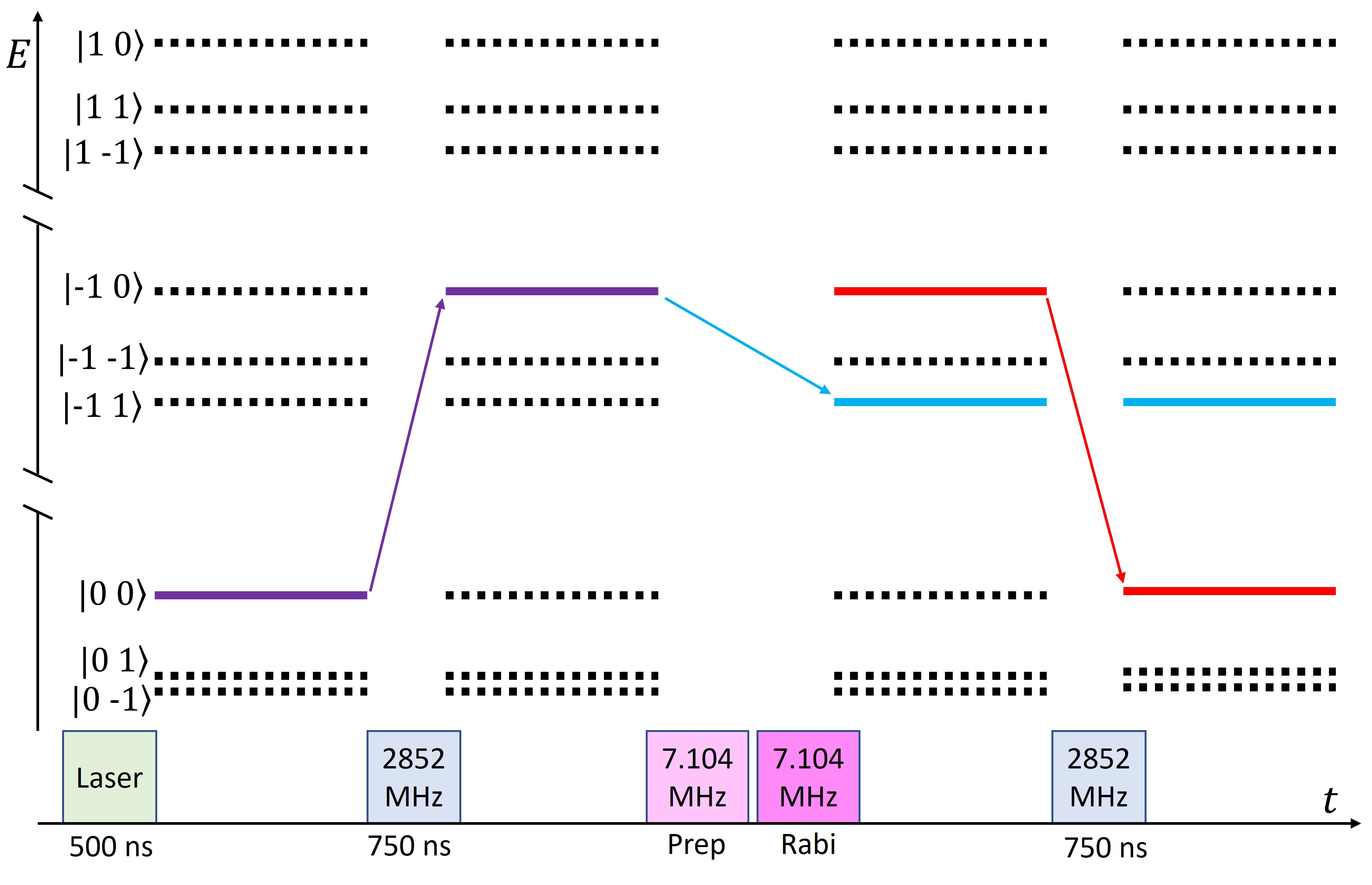}
    \caption{Action of the spin gates of the initialisation (left) and Rabi tomography (right) sequences on NV centre states. The states are written in the format $\ket{m_S,m_I}$. Dashed lines represent empty states, and solid-colored lines represent populated states. Different coloring of the states is to guide the eye. The 7.11 MHz RF pulses in the Rabi tomography sequence are the preparation of an arbitrary state (Prep) and the pulse of variable duration that performs the Rabi cycle (Rabi). The two RF pulses are not phase-matched, but have a well-defined phase difference.}
     \label{nvlev}
\end{figure*}

\subsection{Implementation of RQST on a single nuclear spin using two-qubit gates}\label{experiment}
In addition to the single NV electron spin RQST, we demonstrate the Rabi tomography method on the nuclear spin of an NV centre in detail. As the NV nuclear spin cannot be directly read out optically, it can be probed indireclty from the electron spin ancilla as discussed above. This is because the electron spin and $^{14}$N nuclear spin in the (E-N) NV spin register system are connected through the longitudinal part of the hyperfine coupling defined by the last term of (\ref{Hamiltonian}). This results in a more complex initialisation and tomography protocol, which is depicted in figure \ref{nvlev}.\\
The validity of the tomography method is verified by performing it on a set of known states chosen across the Bloch sphere to capture the performance of an arbitrary nuclear spin quantum state, considering the experimental limitations for verifying performance of the method on an arbitrary state. The nuclear state is prepared in advance, based on \cite{chakraborty2017polarizing}. This state preparation is an auxiliary but a required step. Hence, it is shown in figure \ref{nvlev} for completeness.\\
The first step to state preparation is initialisation to the eigenstate $m_I = 0$. While the electron spin can be initialised to $m_S=0$ using a laser pulse, the nuclear spin does not have such an intrinsic optical initialising mechanism. On the contrary, a sufficiently strong laser pulse can destroy the initialisation of the nuclear spin, making the three states of the nuclear spin triplet equally populated and thus resulting in the mixed state $\ket{\psi} = \frac{1}{\sqrt{3}}(\ket{0, 0} + \ket{0, 1} + \ket{0, -1})$. This is explained in more detail in Supplementary Material \cite{Supp}. The spin initialisation process is realized by transition-selective pulses and a non-unitary transformation implemented by a short laser pulse, optimized for maximum spin polarisation. Figure \ref{nvlev}-left depicts the principle of the sequence operation. A laser pulse polarises the electron spin, two microwave (MW) pulses drive two-electron spin transitions, and then the radio-frequency (RF) pulse drives two nuclear spin transitions, followed by another shorter laser pulse, which initializes the electron spin back to $m_S=0$ again. The duration of the last laser pulse is crucial since it has to preserve the nuclear spin polarisation. MW and RF pulses have to exactly invert the population (a $\pi$-pulse), therefore the optimal duration, as well as the exact frequency of these pulses, also had to be determined in advance (see Supplementary Material \cite{Supp}). The result of the nuclear spin polarisation has been determined experimentally from Lorentzian fits of the resonance peak profiles, as discussed in Supplementary Material, as estimated from the relative depths of the peaks. Hence, the starting state can be determined: $\ket{\psi_0} = \sqrt{0.6543} \ket{0, 0} + \sqrt{0.1484} \ket{0, 1} + \sqrt{0.1973} \ket{0, -1}$ assuming a perfect initialisation of the electronic spin to $m_s = 0$. Possible reasons could be an incomplete laser polarisation of the electron spin or nuclear spin mixing, which happens during the electron spin initialisation.
In the following, we shall work with $\ket{\psi}$ as the logical state $\ket{0}$, which under perfect initialisation would correspond to $\ket{m_s, m_I} = \ket{0, 0}$. \\
Once the nuclear spin is polarised to $\ket{0}$, an arbitrary state needs to be prepared. For that reason, we drive the transition between $m_I=0$ and $m_I=1$ for an arbitrary period of time with the preparation RF pulse. At $m_s=0$ the nuclear levels $m_I=1$ and $m_I=-1$ are not well resolved, so preceding the preparation pulse is first the electron spin transition $m_S=0$ to $m_S=-1$ (figure \ref{nvlev}-right). Once an arbitrary state has been prepared, an x-phase or y-phase RF pulse of varying duration drives the Rabi cycle between $\ket{0}$ and $\ket{1}$. After that, the population of $\ket{-1,0}$ (in the notation $\ket{m_S,m_I}$) is transferred to $\ket{0,0}$ with the MW pulse, where it can be read. The full sequence has to be repeated for each duration of the Rabi pulse to obtain the desired signal.

\section{Experimental Results}
\subsection{Results on single electron spin}
\begin{figure*}[htbp]
    \centering
    \includegraphics[width=\textwidth]{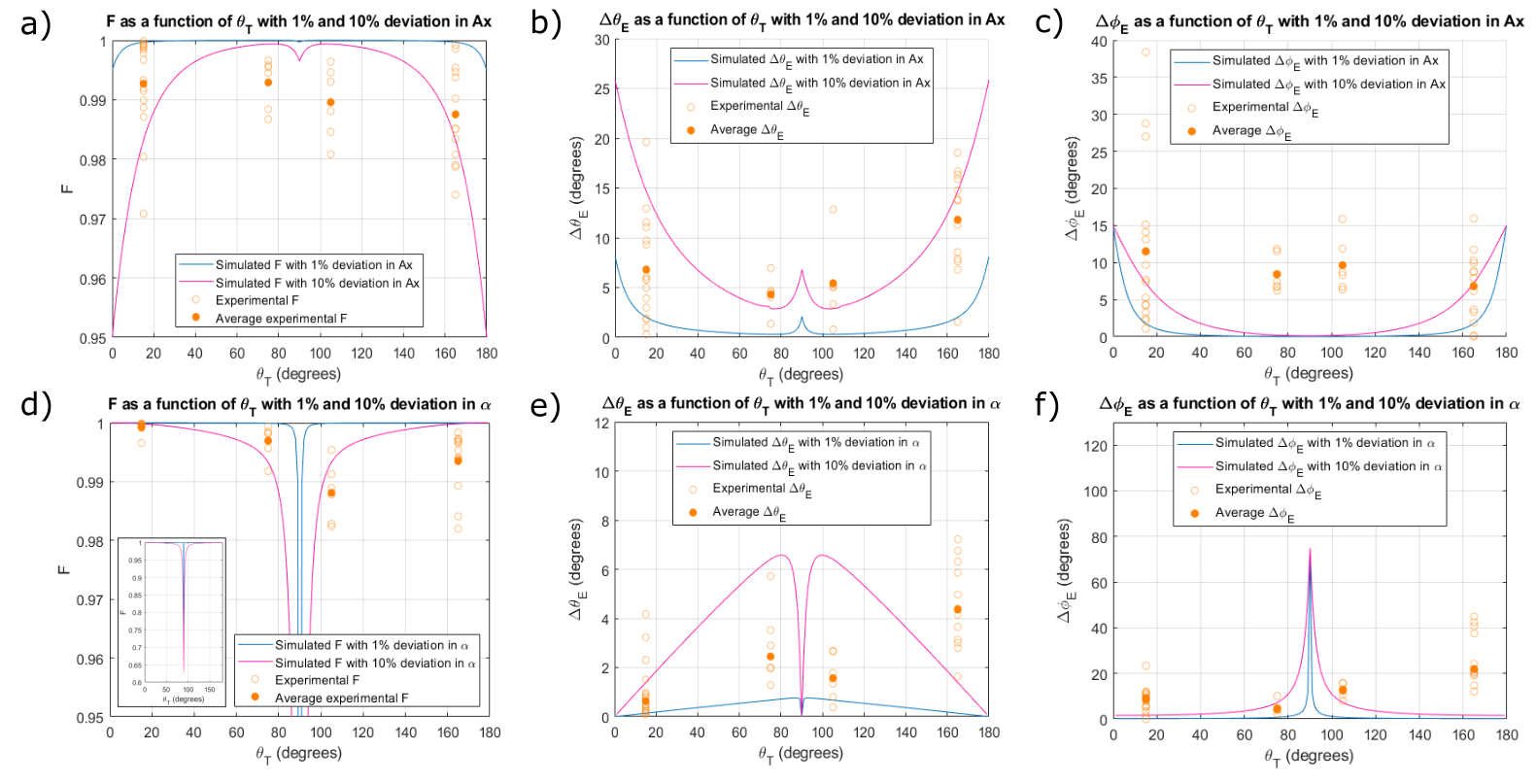}
    \caption{Experimental (orange circles), average experimental (orange dots) and simulated (blue and pink traces) QST results as function of the phase angle $\theta_\text T$ of the prepared state. The blue (pink) simulated curve represents a 1$\%$ (10$\%$) error in the QST parameters $A_x$ for RAQST and $\alpha$ for RPQST. \textbf{(a)} Fidelity, \textbf{(b)} error on $\theta$: $\Delta\theta_E = |\theta_T - \theta_E|$, \textbf{(c)} error on $\phi$: $\Delta\phi_E = |\phi_T - \phi_E|$ for RAQST and \textbf{(d)}, \textbf{(e)}, \textbf{(f)} for RPQST. Here, $(\theta_E, \phi_E)$ is the state that is determined with a $1$ or $10\%$ error in one Rabi parameter ($A_x$ or $\alpha$). The simulated curves indicate that the errors are larger at the poles for RAQST and at the equator for RPQST. wIn (d), the inset shows the full scale of the simulated data. The absence of the simulated trends in the experimental data indicates that the errors on $A_x$ and $\alpha$ are not the dominant reason for the fidelity decrease in our measurements. The notably high fidelity at $\theta_\text T=15^\circ$ in (d) is discussed in Supplementary Material \cite{Supp}.}
    \label{fig:F}
\end{figure*}
To study the performance, the reliability and statistics of the RQST method, we carried out  40 measurements of 9 different states (which are specified below) of the NV electron spin state at room temperature. An average fidelity of (0.991 $\pm$ 0.004) is found for RAQST and (0.995 $\pm$ 0.005) for RPQST (Fig. \ref{fig:F}).   We attribute this difference to our observation that sample drift during the measurements causes the ODMR contrast (and thus Rabi amplitude) to deteriorate.\\
For the 9 states, the phase angles $\theta$ and $\phi$ were varied strategically to show how the fidelity of state determination depends on the state itself. The states that were measured are $(15^{\circ}, 255^{\circ})$, $(15^{\circ}, 225^{\circ})$, $(15^{\circ}, 195^{\circ})$, $(75^{\circ}, 255^{\circ})$, $(75^{\circ}, 225^{\circ})$, $(105^{\circ}, 255^{\circ})$, $(105^{\circ}, 225^{\circ})$, $(165^{\circ}, 255^{\circ})$ and $(165^{\circ}, 225^{\circ})$. As $\theta$ is varied between its full range of $[0^{\circ}, 180^{\circ}]$ and $\phi$ is varied within an octant $[180^{\circ}, 270^{\circ}]$, we cover a range of angles which is representative for the entire Bloch sphere. From both phase angles, $\theta$ showed the largest influence on $F$. 
Based on (\ref{psi}) - (\ref{phh}), one can also calculate how the relation between the fidelity and experimental errors depends on the polar angle $\theta_T$ of the prepared state. We simulate this in Figure \ref{fig:F} by calculating the parameters $A_x$ and $\alpha$ for the case $F=1$ (state determined by QST equal to prepared state) and then calculating $F$, $\Delta \theta_E = |\theta_T - \theta_E|$ and $\Delta \phi_E = |\phi_T - \phi_E|$ ($\theta_E$ and $\phi_E$ being the angles determined from the QST simulation) in the case when $A_x$ and $\alpha$ are changed by $1\%$ or $10\%$ arbitrarily (Fig. \ref{fig:F}-solid lines). From these simulations it follows that the determined state angles ($\theta_E$, $\phi_E$) and fidelity are affected differently by experimental errors depending on $\theta_T$ and on the applied QST methodology, i.e. RAQST or RPQST. The former shows a higher susceptibility to error for $F$ near the poles of the Bloch vector and a lower influence close to the equator, the latter shows the opposite behaviour. The same observations are true when the simulated error is extended to the other Rabi parameters $A_y$ and $\beta$, which is shown in detail in Supplementary Material \cite{Supp}. \\
The experimental data do not fully match the simulated dependence on the polar angle; the errors are consistent across various polar angles. We thus speculate that the origin is in experimental factors that are invariant to the prepared state. Also, we note that for example, small imperfections in phase error can still yield very high fidelities.\\

\subsection{Results on single nuclear spin using two-qubit gates}
\begin{figure}[htbp]
    \centering
    \includegraphics[width=\linewidth]{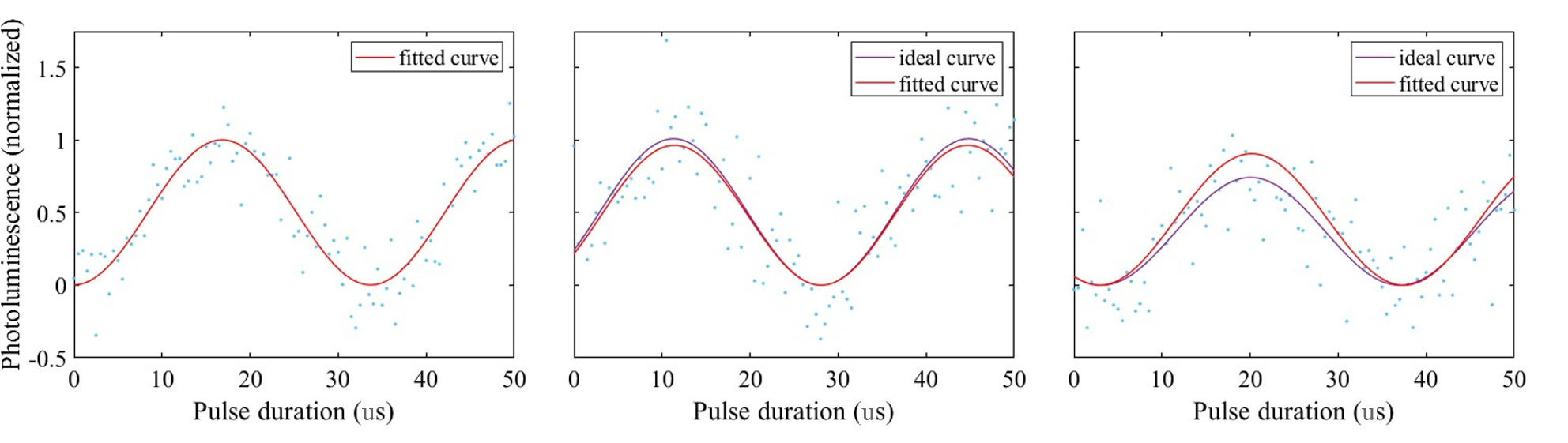}
    \caption{Nuclear Rabi tomography experiment results for the state ($\theta = 58^{\circ}$, $\phi = 249^{\circ}$) (x-Rabi - middle and y-Rabi - right) and the reference measurement (left). Experimental data is fitted (red) and expanded with the Rabi oscillations to be expected (purple) for the ideal state ($\theta = 58^{\circ}$, $\phi = 249^{\circ}$). The amplitudes of the theoretically expected and experimentally measured Rabi oscillations differ to a higher degree than their respective phases, leading to the lower fidelity obtained with RAQST.}
    \label{Rabi1}
\end{figure}

\begin{figure*}[htbp]
    \centering
    \includegraphics[width=\textwidth]{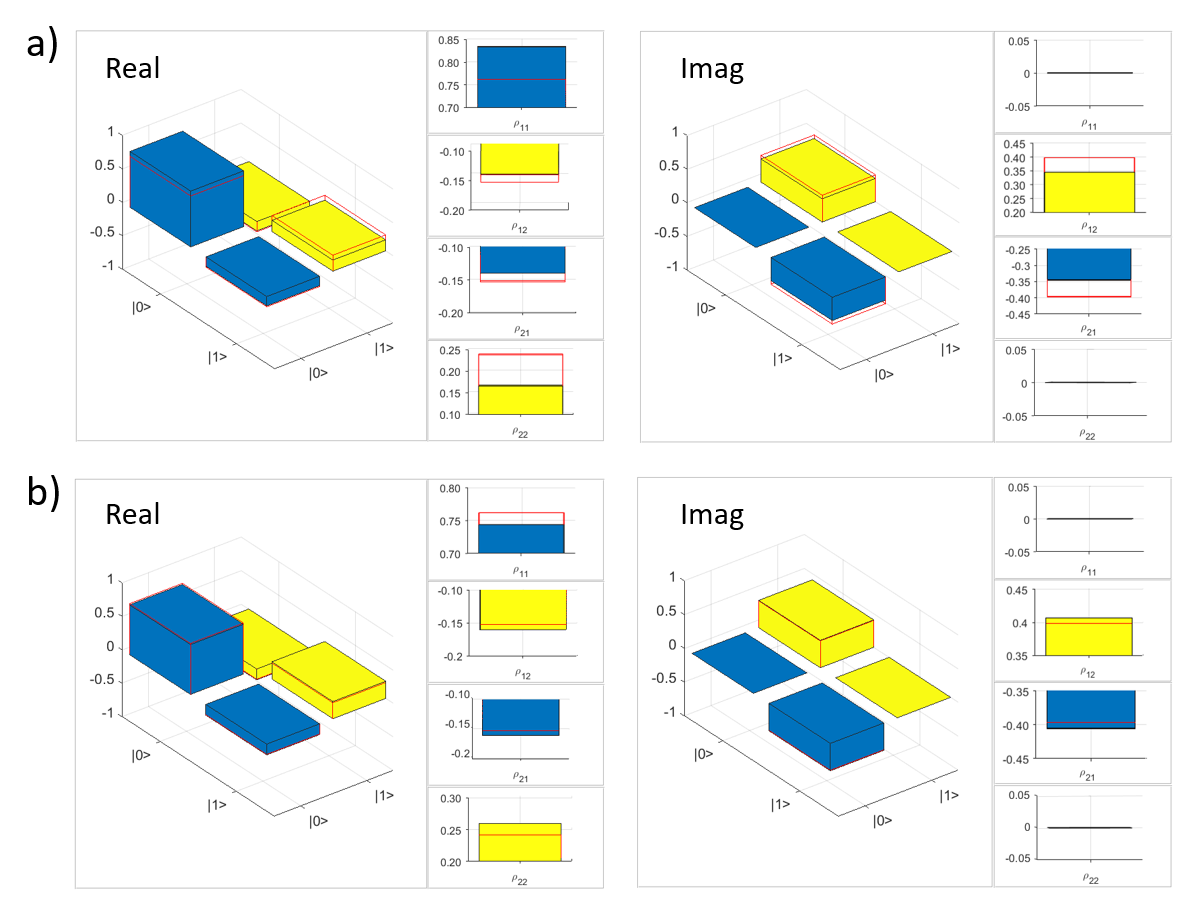}
    \caption{Real and imaginary parts of the experimentally determined density matrices in comparison to the expected density matrices (red overline), for which the corresponding state parameters are ($\theta=58^{\circ}$ and $\phi=249^{\circ}$). (a) Results for RAQST, that give a fidelity of 0.9919. (b) Results for RPQST, that give a fidelity of 0.9995 -our maximum fidelity for nuclear spin tomography. }
    \label{dm}
\end{figure*}

Using the implementation protocols described in Figure \ref{nvlev}, we perform the  Rabi QST  methodology on the NV nuclear spin that is read from the electron spin. We demonstrate the method on three different generic states ($(\theta, \phi) $ = $(58^{\circ}, 249^{\circ})$, $(137^{\circ}, 53^{\circ})$ and $(58^{\circ}, 270^{\circ})$ -each measured twice). These states are chosen at random and lie in different octants of the Bloch sphere. In Figure \ref{Rabi1}, we show an example of raw data and the fits. The fits are referenced to the curves expected for the prepared state, marked "ideal curve". Using the obtained density matrix corresponding to the state $\ket{\psi}$ for $\theta=58^{\circ}$ and $\phi=249^{\circ}$, a fidelity $F_\text{A}$ of $(0.995 \pm 0.005)$ is determined for two repetitions of the presented RAQST measurement and a fidelity of $F_\text{P}$  of $(0.999 \pm 0.0007)$ for RPQST. The measured density matrix for this state is presented in Figure \ref{dm}. For all six measurements of the three different states the average fidelity is (0.99 $\pm$ 0.01) for RAQST and (0.997 $\pm$ 0.003) for RPQST. The value for RAQST is relatively low due to the higher amount of sample drift in the longer nuclear measurements, which was especially detrimental for the state vector near the y-axis. The fidelities for the nuclear spin are of the same order as for the electron spin. This indicates that the operations necessary to prepare nuclear state and read it (Fig. \ref{circuit}) do not introduce a significant error to the measurement.

Finally, in Supplementary Material we provide modelling of the quantum state fidelity as a function of the state mixedness defined as the ratio of the Rabi sphere radius for a partially mixed state and for the pure state. In particular, our intention was to visualise how the state mixing due to decoherence can contribute to the reduction of the QST fidelity \cite{Supp}. In our experiment, the longitudinal coherence of the electron spin is in the range of milliseconds and the nuclear spin in the range seconds, the population spin mixing will not contribute dramatically to the measured fidelity. Also the Rabi traces are modelled using Rabi functionals to mitigate any influence of decoherence. This approach is to some extend similar to the ref \cite{Zhang23}, where re-normalised photon counting was used. The possibility of directly evaluating the Bloch sphere radius by RAQST allows to monitor the decoherence and the state mixing. 

\section*{Conclusion}\label{conclusion}
We have devised a novel method for QST with two variants RAQST and RPQST involving Rabi experiments combined with unitary operations. The average fidelities we obtain on the electron spin are (0.995 $\pm$ 0.0048) with RPQST and (0.991 $\pm$ 0.00363) with RAQST,  from about 40 measurements on states with 9 different final state vector positions on the Bloch sphere. For the nuclear spin, similar fidelities are obtained on three different states with a maximum fidelity of $(0.999 \pm 0.0007)$ for RPQST. RPQST outperforms RAQST as it suppresses systematic experimental errors related to the sample drift. We have shown by modelling that the sensitivity of the method to errors depends on the position of the prepared state-vector on the Bloch sphere, more precisely on $\theta$. For RAQST, states close to the equator are less sensitive to experimental error, compared to states at the poles. The inverse is true for RPQST. Taking this into account, states that are more sensitive to errors might be selected for optimisation of the qubit gates. We have also demonstrated by modelling that small errors on phase or amplitude do not necessarily deteriorate the fidelity significantly. Further on, the modelling of the fidelity in case of white noise has been carried out, showing that our method can outperform the QST from projective measurements for certain ranges of states defined on the Bloch sphere. We claim that our technique will be useful for widening of QST methodologies for the situations with a increasing number of qubits related to obtaining the complete set of the projections which prolongs exponentially the measurement time. This  can be especially useful for systems where the spin state are not directly observable, such as solid state spin systems. In this case we replace the unitaries by direct Rabi measurements.\\
Though our spin control already reaches high fidelity numbers compared to state-of-the-art, further enhancement of fidelity is possible through characterising the systematic errors and suppressing the decoherence effect by, for example, using dynamical decoupling pulses or numerically designed pulses using optimal control theory to get rid of systematic errors in the implementation. In addition, improving the hardware for suppressing inhomogeneity in the microwave control field would also have a positive effect on the fidelities we obtain.\\
Though, we have demonstrated our new method for QST using ODMR, the work on photoelectric detection of magnetic resonance (PDMR)\cite{Bourgeois} QST is an ongoing venture. The PDMR-QST will be an unavoidable tool for achieving high-fidelity state engineering in a scalable dipole-dipole coupled NV spin register desirable for fault-tolerant programmable quantum computers.\\
We hope that our work contributes towards placing NV centre-based solid state spin-qubit systems among the prestigious group of error correction-viable quantum computing architectures.
\section{Acknowledgement}
We acknowledge the support from Eu projects 101113901 and 101113983 (QU-Test and QU-Pilot projects), Grant Agreement No. 101135699 (SPINUS), Grant Agreement No. 101135359 (C-QUENS), GACR project 24-12984S, FWO Project No. G0D1721N and EOS CHEQs Project No. 40007526. Research was co-financed by the Research Foundation-Flanders via FWO file number 11O1625N.
\bibliography{ref2}
\bibliographystyle{apsrev4-2}
\end{document}